\documentclass[11pt]{article}
\usepackage[latin1]{inputenc}
\usepackage{geometry}
\usepackage{amssymb,amsmath}
\usepackage{graphicx,color}
\usepackage[english]{babel}
\usepackage{mathrsfs}

\newcommand{\ave}[1]{\ensuremath{\left \langle {#1} \right \rangle}} 
\newcommand{\spectral}{\ensuremath{\mathscr{S}}} 

\newcommand{\de}{\ensuremath{\delta}}

\newcommand{\om}{\ensuremath{\omega}}
\newcommand{\Om}{\ensuremath{\Omega}}

\graphicspath{{figures/}} 
\DeclareGraphicsExtensions{.pdf,.png,.jpg} 

\begin{document}
\begin{titlepage} 
\title{Grand-mother clocks and quiet lasers.}

\author{%
Jacques \textsc{Arnaud}
\thanks{Mas Liron, F30440 Saint Martial, France},
Laurent \textsc{Chusseau}
\thanks{Institut d'\'Electronique du Sud, UMR n°5214 au CNRS, Universit\'e Montpellier II, F34095 Montpellier, France},
Fabrice \textsc{Philippe}
\thanks{LIRMM, UMR n°5506 au CNRS, 161 rue Ada, F34392 Montpellier, France}
}
\maketitle

\begin{abstract}      

Galileo noted in the 16\textsuperscript{th} century that the period of oscillation of a pendulum is almost independent of the amplitude. However, such a pendulum is damped by air friction. The latter may be viewed as resulting from air molecules getting in contact with the pendulum. It follows that air friction, not only damps the oscillation, but also introduces randomness. In the so-called ``grand-mother'' clock, discovered by Huygens in the 18\textsuperscript{th} century, damping is compensated for, on the average, by an escapement mechanism driven by a falling weight. The purpose of this paper is to show that such a clock is, in its idealized form, a quiet oscillator. By ``quiet'' we mean that in spite of the randomness introduced by damping, the dissipated power (viewed as the oscillator output) does not fluctuate slowly. Comparison is made with quiet laser oscillators discovered theoretically in 1984. Because the input power does not fluctuate in both the mechanical oscillator and the quiet laser oscillator, the output power does not fluctuate at small Fourier frequencies, irrespectively of the detailed mechanisms involved.

\end{abstract}

\end{titlepage}

\newpage

\section{Introduction}\label{introduction}

Oscillators are devices whose mass position or electrical potential vary sinusoidally as a function of time. The time period $T$ is a constant in the present paper and we are not concerned with frequency fluctuations. Only small amplitude fluctuations about some large oscillation amplitudes are considered, first for the ``grand-mother clock'', and later for laser oscillators. 

In any practical oscillator there is damping. Accordingly, in order to sustain the oscillation some energy must be continuously fed in. In the case of the grand-mother pendulum, the energy originates from a slowly dropping weight. In the case of laser diodes, the energy originates from an electrical current, or some other source of energy such as thermal radiation. These power supplies are then called ``pumps''. The amplitude fluctuations we referred to above are primarily caused by damping, which should be viewed as a random process. Damping may be caused for example by air molecules contacting the pendulum at random times. What we consider as a measurable quantity is not the oscillation amplitude itself, but the dissipated power $P(t)$. Note that the oscillators treated here involve a single input channel (through which the power is fed in), and a single output channel corresponding to the dissipated power $P(t)$. All losses aside from those resulting from the damping mechanism are supposed to be negligible, and the oscillator is stationary, which means that a shift in time would not affect the system operation.
 
The main question we wish to consider is the following: Assuming that the power fed into the oscillating mechanism is strictly constant in time, does the dissipated power $P(t)$ fluctuate? Under ideal conditions and for slow variations (small Fourier frequencies), our answer is that the dissipated power does not fluctuate. The general argument is as follows. In a conservative device (i.e., with no internal loss or gain) the output power may fluctuate even if the input power is a constant. This is because mechanical or electrical energy is stored in the device. However, the law of conservation of energy entails that the output power, integrated over a sufficiently long time, must be equal to the input power, integrated over the same long time interval\footnote{As an example, note that in the electrical power grid where energy storage is negligible, the power generated by, say, nuclear plants must, at any instant, be equal to the consumed power, or nearly so. Because the consumed power varies quickly in time but the power generated by nuclear plants cannot be made to vary quickly, burning-gas generators are required from time to time. Of course, the law of conservation of energy tells us nothing about the power distribution among various customers.}. In the case of the grand-mother pendulum, the input power, fed in with the help of an escapement mechanism, is a constant. We prove in the next section, both analytically and through numerical calculations, that, in agreement with the above general argument, the dissipated power $P(t)$ does not fluctuate at small Fourier frequencies. Here damping events occur at times that are Poisson distributed (crudely speaking this means that the damping events are ``random'' in time). But some of these events absorb a large energy while others absorb a small energy. This is why the absorbed power may be regulated. 

A laser diode, on the other hand, may be fed in by a constant current, generated for example by a battery in series with a large cold resistance. The electrical potentials across the laser diode and across the detector diode that convert the emitted light back into an electrical current, are nearly equal to $h/Te$, where $h$ denotes the Planck constant, $T$ the oscillation period, and $e$ the electron charge. It follows that if a constant current is fed into a laser diode, conservation of energy alone tells us that the photo-detector, which is the only cause of loss according to our assumptions, delivers a current that does not fluctuate either at small Fourier frequencies. In the present situation, each photo-detection event carries the same energy because the electron charge is a constant. It follows that regulation of $P(t)$ requires that the detector electronic events be distributed in time in a ``sub-Poissonian'' manner. By that, one means that the event occurrence times are more regularly spaced than in a Poissonian distribution. 

It is often argued that nonclassical features of the electromagnetic field states such as sub-Poissonian statistics are consequences of the field quantization. For sure Quantum Optics provides a framework for a complete interpretation of nonclassical light features. The first theory of ``sub-Poissonian'' emission by regularly driven laser diodes was given that way \cite{golubev:JETP84}. But we show here that accounting for the emitter-detector system as a whole quantum device produces the same interpretation only using  the law of conservation of energy. The formulas obtained  for the photodetection statistics thus agree in every details with the results obtained independently by quantum theorists. Some other measurable effects do indeed require light quantization. However, this is not the case for the kind of oscillators that we are considering. Our view point is that lasers, on that respect, are akin any oscillator.

\section{The ``grand-mother'' clock}\label{clock}

The basic element of a grand-mother clock is a weight $W$ suspended at the end of a weightless bar of length $L$ in the earth gravitational field $g$. As was first shown by Galileo the oscillation period $T=2\pi \sqrt{L/g}$ does not depend on the oscillation amplitude as long as this amplitude remains small, a condition that we assume fulfilled. The period $T$ does not depend either on the weight value according to the equivalence principle: inertial mass equals gravitational mass. For simplicity we suppose that the pendulum period is unity, that is $T=1$~s. This amounts to selecting some appropriate $L$ value, considering that $g\approx10$ m/s$^2$. We also suppose that $W=1$ so that  the highest weight altitude $E$ represents the pendulum energy since the kinetic energy then vanishes. In the following we denote by $E_k, ~k=1,2,...$ the pendulum energies at successive periods of oscillation. This energy gets decremented by a random damping mechanism to be specified below, and incremented by a regular escapement mechanism.

\begin{figure}
\centering
\includegraphics[scale=1]{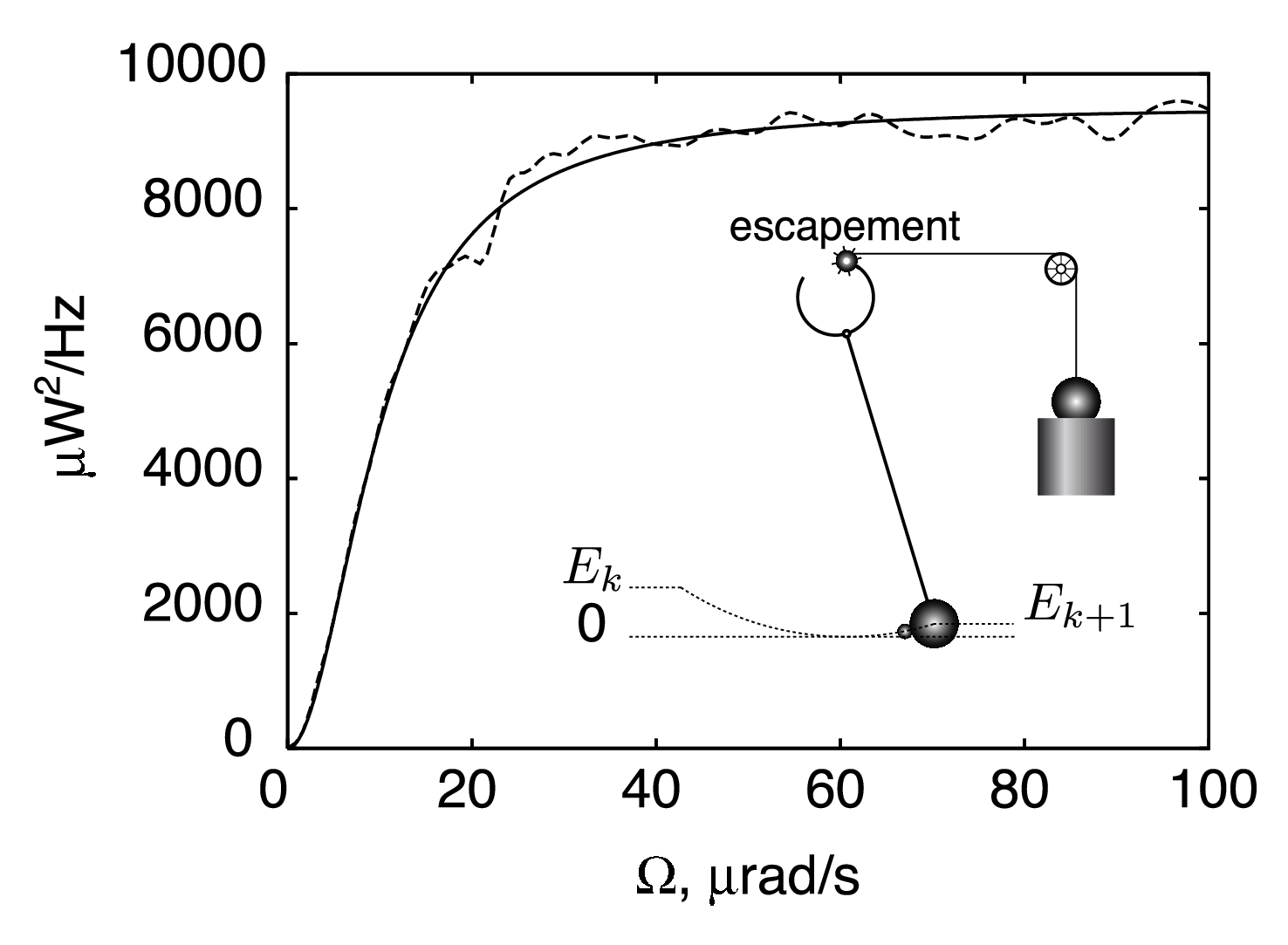}
\caption{The figure represents the ``grand-mother'' clock, discovered by Huygens in the 18\textsuperscript{th} century. The pendulum consists of a weight at the end of a weightless bar. In our model, damping is caused by molecules of weight $w$ being raised by the pendulum from the lowest to the highest weight level, with probability $p\ll1$. Damping is compensated for, on the average, by an escapement mechanism driven by a falling weight delivering a constant energy $\delta$ per period. The curves show the spectral density of the dissipated power $P(t)$ as a function of the Fourier frequency $\Omega/2\pi$, numerically evaluated over 10$^8$ periods of oscillation (irregular curve) and obtained analytically (smooth curve). This clock, in its idealized form, is a quiet oscillator in the sense that in spite of the randomness introduced by damping the dissipated power does not fluctuate at small Fourier frequencies.}
\label{fig}
\end{figure}

Let us first describe the damping mechanism. The pendulum, with energy $E_k$, is supposed to pick up with probability $p\ll1$ at each period a molecule of weight $w$ at rest at the lowest level, and to release it at the highest level $E_{k+1}$ (see Fig.~\ref{fig}). Because the probability $p\ll1$, the molecule-picking events form a Poisson process. The average inter-event time for a Poisson process is known to be $1/p$ \cite{Papoulis,gillespie}. Note that we are considering only time intervals much larger than the pendulum period $T=1$. Raising a molecule of weight $w$ from altitude $0$ to altitude $E_{k+1}$ amounts to reducing the pendulum energy from $E_k$ to $E_{k+1}=E_k-wE_{k+1}$ according to the law of energy conservation. It follows that, if a molecule-picking event occurs (a rare event), we have: $E_{k+1}=E_k/(1+w)\approx (1-w)E_k,~w\ll1$. 

In order to maintain a constant oscillation amplitude, at least on the average, a power supply is required. An adequate mechanism was proposed in the 18\textsuperscript{th} century by Huygens, who invented what we call here the grand-mother clock. Power is delivered by a weight suspended at the end of a cord. An escapement mechanism (crudely represented in the figure) allows the suspended weight to drop by a fixed height at each swing of the pendulum, thereby delivering to it a constant energy (or power since $T=1$) that we denote $\delta$. The pendulum average energy $\ave{E}$ is obtained by equating the input power $\delta$ and the average absorbed power $pw\ave{E}$. Thus, $\ave{E}=\delta/pw$. It is appropriate in numerical calculations to begin with an initial pendulum energy equal to $\ave{E}$. At every period ($k=1,2,...$) we add to the pendulum energy the energy delivered by the escapement mechanism, that is $E_{k+1}=E_{k}+\delta$. We also select a random number $x$ uniformly distributed between 0 and 1. If $x<p$ (a rare event), we subtract from the pendulum energy the molecule-raising energy: $E_{k+1}=(1-w)E_{k}$.

While the power supply is constant in time according to the above discussion, the damping mechanism has a random character. Our purpose is to evaluate the energy released by the molecules as a function of time. As said above, this energy is generated at times (called ``events'') corresponding to a Poisson process. But the energy released by the molecules varies from event to event. If a molecule-picking event occurs at time $k_i$ and the next one occurs after an anomalously-large time ($k_{i+1}-k_i\gg1/p$), the pendulum energy has been much incremented and therefore the next event absorbs a larger-than-usual energy. This is how one can explain in a qualitative manner the mechanism behind dissipation regulation. As the figure shows, the spectral density of this so-called ``marked'' Poisson process \cite{Cox1980} has been found numerically to be in excellent agreement with the analytical formula given below, which has been obtained as in \cite{Arnaud1995} for laser diodes.
\begin{align}
\label{intro}
\spectral(\Om)=\frac{\delta^2/p}{1+(pw/\Om)^2}.
\end{align}
The spectral density vanishes as the Fourier angular frequency $\Om\to 0$. This must be the case because, as we discussed earlier, for slow variations, the conservation of energy implies the conservation of power, the stored energy being then negligible. In our numerical application, $w=1$~mN, $\de=10\;\mu$J, $p=0.01$ and thus the average inter-event time is 100~s. 
 
\section{Lasers}\label{lasers}

Let us now compare grand-mother clocks to lasers driven by non-fluctuating pumps, the pendulum angular frequency $2\pi/T$ corresponding to the laser angular frequency $\om$. In the case of the grand-mother clock the dissipation event times are Poisson distributed, but each event has a mark that varies so that a regulation mechanism may occur. In the case of quiet lasers, that is, lasers driven by a non-fluctuating source of energy, the photo-detection events have all the same energy. This energy is $\hbar\omega$ (where $\hbar$ denotes the Planck constant divided by $2\pi$) because the electron charge $e$ is a constant and the electric potential across the detector is approximately $\hbar\omega/e$. Output power regulation in that case follows from the fact that the event times are sub-Poissonian. That is, the time interval between successive events is nearly constant. 

Many experimental results in Physics may be explained only if one treats the field in an optical cavity in analogy with the motion of quantized harmonic oscillators. This procedure is referred to as a ``Quantum Optics'' treatment. This universal framework may induce some doubt that the quietness of constant current driven lasers can be so simply derived from the law of energy conservation. Let us now explain by way of examples rather than in a formal manner why, nevertheless, the results concerning oscillator noise we are interested in are unaffected by light quantization.

Let us consider a thought experiment with an optical cavity at a temperature of 0~K. At such temperature the probability that the cavity contains optical photons is null. According to the Quantum-Optics terminology, the cavity field is then in the ``vacuum state'', which corresponds to the ground state of an harmonic oscillator. Suppose next that a resonant two-level atom in the excited state is injected into the optical cavity at a prescribed speed.
If this speed is appropriate, the Quantum Optics treatment predicts that the atom exiting the cavity is, with certainty, in the ground state, a prediction which has been verified with fair accuracy by the Haroche-group experiments  that clearly show for the first time the quantum Rabi oscillation of a single atom in a singlemode cavity \cite{PhysRevLett.76.1800}.
In contradistinction, according to the semi-classical theory, the classical Rabi equations then predict that the atom exits from the cavity in the excited state: No interaction with the cavity would have taken place. Clearly the outcome of such experiments may be explained only through quantization of the light field and rely on the Jaynes-Cummings Hamiltonian formalism \cite{Jaynes1963}.

But let us go a bit further with the thought experiment. Suppose that a similar atom, this time in the ground state, is subsequently injected in the cavity. It is then observed that this second atom leaves the cavity in the excited state\footnote{It is of course assumed that the cavity losses are very low, so that the cavity field did not significantly decay at the second atom arrival time.}. If we consider the two atoms together, one may say that the energy of the exiting atoms is the same that the energy of the entering atoms, namely $\hbar\omega$. Energy is thus conserved on the average. But light quantization implies that for some period of time the cavity contains energy. If a stream of atoms alternately in the excited state and in the ground state is sent into the cavity, the above discussion shows that the output atomic stream is very much like the input atomic stream except for a small time delay. 

Let us recall that the oscillators we are considering possess only one input (the ``pump'') and one output (the photo-detector). If we separate spatially the stream of atoms initially in the excited state and the stream of atoms initially in the ground state, the Quantum-Optics treatment makes a prediction drastically different from the semi-classical treatment. But this is not allowed in our model. Obviously, the law of conservation of energy alone cannot tell us how much energy goes into different ports. 
Finally, let us comment on the ``small time delay'' mentioned above, which follows from the Quantum Optics treatment, but not from the semi-classical treatment. One should recall that our laser-oscillator noise theory applies only to \emph{stationary} systems, which are unaffected by time lags. This delay therefore needs not be considered.

From 1989 up to this time, we repeatedly published predictions of sub-Poissonian photo-electron statistics on the basis of this semi-classical theory. The explicit expressions obtained for the photo-electron statistics have always been found to coincide exactly with the available Quantum-Theory results. On the one hand, a tutorial presentation of the concepts involved in the semi-classical theory, with all unnecessary details being removed, were given in Refs.~\cite{Arnaud1995, Arnaud2002}. On the other hand,  Ref.~\cite[see Eq.~(17)]{Chusseau2002b} relates to the sub-Poissonian output of four-level lasers. The levels are labelled by 0, 1, 2, 3, levels 1-2 being the lasing levels. This result, which demonstrates sub-Poissonian light statistics and involves four independent parameters relating to the various spontaneous decay rates and pumping rates, is exactly the same as the one obtained from Quantum Optics in Ref.~\cite[see Eq. (4)]{Ritsch1991}. In four-level lasers sub-Poissonian light may be observed even when the pumping light is in the coherent state (a coherent light incident on a photo-detector generates Poissonian photo-electrons). The regulation of the output light statistics in that case originates from the spontaneous decays from levels 3 to 2, and from levels 1 to 0 that play a role somewhat similar to the large cold resistor employed in quiet laser diodes.

\section{Conclusion}\label{conclusion}

We considered in the present paper a classical oscillator, the ``grand-mother clock'', discovered by Huygens in 18\textsuperscript{th} century, which may be called a quiet oscillator because the dissipated power does not fluctuate slowly. This is a consequence of the fact that the escapement mechanism delivers a constant power. In the case of quiet lasers the dissipated power does not fluctuate slowly either because the input power (i.e., the pump) is a constant. In that case each event possesses the same energy, but the occurrence times are sub-Poissonian, i.e., regularly spaced. We hope to have convinced the reader that the law of energy conservation suffices to understand why some classical or quantum oscillators may be ``quiet''.

\bibliographystyle{IEEEtran}
\bibliography{circuittheory}

\begin{thebibliography}{10}
\providecommand{\url}[1]{#1}
\csname url@samestyle\endcsname
\providecommand{\newblock}{\relax}
\providecommand{\bibinfo}[2]{#2}
\providecommand{\BIBentrySTDinterwordspacing}{\spaceskip=0pt\relax}
\providecommand{\BIBentryALTinterwordstretchfactor}{4}
\providecommand{\BIBentryALTinterwordspacing}{\spaceskip=\fontdimen2\font plus
\BIBentryALTinterwordstretchfactor\fontdimen3\font minus
  \fontdimen4\font\relax}
\providecommand{\BIBforeignlanguage}[2]{{%
\expandafter\ifx\csname l@#1\endcsname\relax
\typeout{** WARNING: IEEEtran.bst: No hyphenation pattern has been}%
\typeout{** loaded for the language `#1'. Using the pattern for}%
\typeout{** the default language instead.}%
\else
\language=\csname l@#1\endcsname
\fi
#2}}
\providecommand{\BIBdecl}{\relax}
\BIBdecl

\bibitem{golubev:JETP84}
Y.~M. Golubev and I.~V. Sokolov, ``Photon antibunching in a coherent light
  source and suppression of the photorecording noise,'' \emph{Sov. Phys.-JETP},
  vol.~60, pp. 234--238, 1984.

\bibitem{Papoulis}
A.~Papoulis, \emph{{P}robability, {R}andom {V}ariables, and {S}tochastic
  {P}rocesses}, 3rd~ed.\hskip 1em plus 0.5em minus 0.4em\relax New York:
  MacGraw-Hill, 1991.

\bibitem{gillespie}
D.~T. Gillespie, \emph{{M}arkov {P}rocesses: {A}n {I}ntroduction for {P}hysical
  {S}cientists}.\hskip 1em plus 0.5em minus 0.4em\relax San Diego: Academic
  Press, 1992.

\bibitem{Cox1980}
D.~R. Cox and V.~Isham, \emph{Point processes}.\hskip 1em plus 0.5em minus
  0.4em\relax London: Chapman, 1980.

\bibitem{Arnaud1995}
J.~Arnaud, ``Classical theory of laser noise,'' \emph{Opt. Quantum Electron.},
  vol.~27, pp. 63--89, 1995.

\bibitem{PhysRevLett.76.1800}
M.~Brune, F.~Schmidt-Kaler, A.~Maali, J.~Dreyer, E.~Hagley, J.~M. Raimond, and
  S.~Haroche, ``Quantum {R}abi oscillation: A direct test of field quantization
  in a cavity,'' \emph{Phys. Rev. Lett.}, vol.~76, no.~11, pp. 1800--1803, Mar
  1996.

\bibitem{Jaynes1963}
E.~T. Jaynes and F.~W. Cummings, ``Comparison of quantum and semiclassical
  radiation theories with application to the beam maser,'' \emph{Proc. IEEE},
  vol.~51, pp. 89--109, 1963.

\bibitem{Arnaud2002}
J.~Arnaud, ``Rate equation theory of sub-{P}oissonian laser light,'' \emph{Opt.
  Quantum Electron.}, vol.~34, pp. 393--410, 2002.

\bibitem{Chusseau2002b}
L.~Chusseau, J.~Arnaud, and F.~Philippe, ``Rate-equation approach to atomic
  laser light statistics,'' \emph{Phys. Rev. A}, vol.~66, p. 053818, 2002,
  quant-ph/0203029.

\bibitem{Ritsch1991}
H.~Ritsch, P.~Zoller, C.~W. Gardiner, and D.~F. Walls, ``Sub-{P}oissonian laser
  light by dynamic pump-noise suppression,'' \emph{Phys. Rev. A}, vol.~44, pp.
  3361--3364, 1991.

\end{thebibliography}

\end{document}